\documentclass[aps,prb,reprint,amssymb]{revtex4-2}

\usepackage{amsmath,amssymb,color,braket,enumitem}
\usepackage{bm}
\usepackage{hyperref}
\usepackage{graphicx}

\DeclareMathOperator{\csch}{csch}

\begin{document}

\title{Phase diagram of  the $J_1$-$J_2$ quantum Heisenberg model for arbitrary spin}

\author{Andreas R\"{u}ckriegel}
\affiliation{Institut f\"{u}r Theoretische Physik, Universit\"{a}t Frankfurt,  Max-von-Laue Stra{\ss}e 1, 60438 Frankfurt, Germany}

\author{Dmytro Tarasevych}
\affiliation{Institut f\"{u}r Theoretische Physik, Universit\"{a}t Frankfurt,  Max-von-Laue Stra{\ss}e 1, 60438 Frankfurt, Germany}

\author{Peter Kopietz}
\affiliation{Institut f\"{u}r Theoretische Physik, Universit\"{a}t Frankfurt,  Max-von-Laue Stra{\ss}e 1, 60438 Frankfurt, Germany}

\date{May 6, 2024}

\begin{abstract}
We use the spin functional renormalization group to investigate 
the $J_1$-$J_2$ quantum Heisenberg model on a square lattice.
By incorporating sum rules associated with the fixed length of the spin operators
as well as the nontrivial quantum dynamics implied by the spin algebra,
we are able to compute the ground state 
phase diagram for arbitrary spin $S$,
including the quantum paramagnetic phase at strong frustration.
Our prediction for the extent of this paramagnetic region for $ S = 1/2 $ agrees well with other approaches
that are computationally more expensive.
We find that the quantum paramagnetic phase disappears for $ S \gtrsim 5 $ due to the suppression of quantum fluctuations with increasing $S$.
\end{abstract}

\maketitle

\section{Introduction}

One of the major challenges in contemporary condensed matter physics 
is the calculation of ground state properties of frustrated spin systems \cite{Balents2010, Sandvik2010, Savary2017, Zhou2017, Diep2020}.
In these systems,
quantum fluctuations play a dominant role at low temperatures and 
may even be strong enough to melt any classical long-range order.
This results in novel, highly entangled quantum paramagnetic ground states like 
spin liquids or resonating valence bond states.
Due to the absence of classical magnetic order,
such quantum paramagnets cannot be investigated using
standard approaches to quantum spin systems such as spin-wave theory. 
At the same time,
sign problems prevent large-scale unbiased quantum Monte Carlo simulations 
for frustrated spin systems \cite{Sandvik2010}.
While the density-matrix renormalization group \cite{White1992, Schollwoeck2005}
offers an efficient means of determining ground state properties of strongly correlated systems in reduced dimensions,  
its computational complexity increases drastically with system size in higher dimensions because of area-law entanglement growth.
Therefore,
there is a need to develop unbiased methods to address frustrated spin systems in arbitrary dimensions.
One such method is the functional renormalization group (FRG) \cite{Wetterich1993, Berges2002, Pawlowski2007, Kopietz2010, Metzner2012, Dupuis2021}.
In a seminal work by Reuther and W\"{o}lfle \cite{Reuther2010},
the established machinery of FRG for electronic systems was applied to a frustrated quantum spin system by representing the spin operators in terms of auxiliary fermions.
Since then,
this so-called pseudo-fermion FRG has been used as an unbiased numerical tool to study a variety of frustrated quantum spin systems in two and three dimensions;
for a recent review, see Ref.~[\onlinecite{Mueller2024}] and references therein.
However,
because of the mapping to auxiliary fermions,
this approach suffers from several disadvantages.
For example, 
the Hilbert space of the fermions contains unphysical states which may contaminate physical results.
This problem has been addressed recently by using either 
a pseudo-Majorana fermion representation of the spin operators \cite{Martin1959, Tsvelik1992, Niggemann2021, Niggemann2022}
or implementing the so-called Popov-Fedotov trick \cite{Popov1988, Schneider2022}.
Moreover, 
since a spin operator is represented by two fermionic operators,
computing $n$-spin correlation functions requires $2n$-fermion correlation functions.
This makes it difficult to include correlations involving more than two spins.
Finally,
because of the nontrivial spin dynamics,
obtaining explicit solutions necessitates heavy numerical computations even for modest truncations of the FRG flow equations.
An alternative FRG approach
to quantum spin systems is the spin FRG,
proposed by Krieg and Kopietz \cite{Krieg2019},
and further developed in Refs.~\cite{Tarasevych2018, Goll2019, Goll2020, Tarasevych2021, Tarasevych2022a, Rueckriegel2022, Tarasevych2022b}.
It relies on a formally exact renormalization group flow equation for the generating functional of the physical spin correlation functions,
without the need to introduce auxiliary boson or fermion operators and their associated subtleties and restrictions.

In our recent  work [\onlinecite{Tarasevych2022b}],
we have demonstrated that even simple truncations of the spin FRG flow 
already yield quantitatively accurate results
for classical, finite temperature phase transitions in quantum spin systems,
for arbitrary spin $S$.
Here, we extend this approach to quantum phase transitions in frustrated spin systems and demonstrate that
the spin FRG is a powerful yet computationally rather cheap 
method to study the zero-temperature phase diagram of quantum paramagnets.
To that end,
we consider a paradigmatic model system for a frustrated quantum magnet:
the antiferromagnetic $J_1$-$J_2$ model with Hamiltonian
\begin{equation} \label{eq:J1-J2}
{\cal H} 
= J_1 \sum_{ \langle i j \rangle_1 } \bm{S}_i \cdot \bm{S}_j
+ J_2 \sum_{ \langle i j \rangle_2 } \bm{S}_i \cdot \bm{S}_j \; 
\end{equation}
on the square lattice,
where $ J_1$, $J_2 \ge 0 $ are antiferromagnetic exchange couplings, 
the spin-$S$ operators $ \bm{S}_i $ satisfy $\bm{S}_i^2 = S ( S+1)$, 
and $ \langle i j \rangle_{ 1 (2) } $
denotes all pairs of nearest (next-nearest) neighbors.
The Fourier transform of the exchange interaction is then
\begin{equation}
J ( \bm{k} ) = 
4 J_1 \gamma_1 ( \bm{k} ) +
4 J_2 \gamma_2 ( \bm{k} ) \; ,
\end{equation}
with the nearest and next-nearest neighbor form factors,
\begin{subequations}
\begin{align}
\gamma_1 ( \bm{k} ) = & \frac{ 1 }{ 2 } \left( \cos k^x + \cos k^y \right) \; , \\[.1cm]
\gamma_2 ( \bm{k} ) = & \cos k^x \cos k^y \; ,
\end{align}
\end{subequations}
where wave vectors are measured in units of the inverse lattice spacing.
This model represents one of the simplest quantum spin systems
where one can study how frustration induces large quantum fluctuations
that destroy classical order.
As such,
it has been intensively studied in the last decades
using a variety of methods,
ranging from spin-wave theory \cite{Chandra1988, Gochev1994, Dotsenko1994}
and Green's function methods \cite{Barabanov1993, Barabanov1994, Siurakshina2001, Sasamoto2024} over large-$N$ expansion \cite{Read1991},
an effective nonlinear $\sigma$ model \cite{Einarsson1991},
Schwinger-boson mean field theory \cite{Einarsson1992},
bond operators \cite{Zhitomirksy1996, Kotov1999, Kotov2000, Doretto2014},
series expansions \cite{Gelfand1989, Oitmaa1996, Sushkov2001, Sirker2006},
exact diagonalization \cite{Dagotto1989, Figueirido1990, Capriotti2000}
density matrix renormalization group \cite{Jongh2000, Jiang2012, Gong2014, Wang2018, Qian2023}, 
Monte Carlo \cite{Capriotti2000, Jongh2000}, 
coupled cluster simulations \cite{Darradi2008},
hierarchical mean-field theory \cite{Isaev2009},
variational approaches \cite{Oliveira1991, Capriotti2001, Murg2009, Zhang2013, Hu2013, Ferrari2020},
pseudofermion FRG \cite{Reuther2010},
as well as neural networks \cite{Nomura2021}.
Apart from the spin-wave calculations,
all of these works focus on the $S=1/2$ case,
where quantum effects are most pronounced.
The consensus from these studies on the phase diagram for $S=1/2$ is as follows:
For weak next-nearest neighbor coupling, $ J_2 \lesssim 0.4 J_1 $, 
the system exhibits 2-sublattice N\'eel order,
with ordering wave vector $ \bm{Q} = ( \pi , \pi ) $.
In the opposite limit of strong next-nearest neighbor coupling, $ J_2 \gtrsim 0.65 J_1 $,
the system orders into a stripe state with ordering wave vector either 
$ \bm{Q} = ( \pi , 0 ) $ or $ ( 0 , \pi ) $.
In the intermediate, strongly frustrated range of exchange couplings,
the ground state is a quantum paramagnet.
The exact nature of this paramagnetic phase,
as well as of the phase transition into it, is still under debate \cite{Figueirido1990, Gelfand1989, Oliveira1991, Read1991, Zhitomirksy1996, Jongh2000, Kotov2000, Sushkov2001, Isaev2009, Jiang2012, Hu2013, Gong2014, Wang2018, Nomura2021, Qian2023}.
It is also not certain \cite{Sandvik2010} whether this exotic phase survives for $S > 1/2$,
as indicated by spin wave theory \cite{Chandra1988}.
As quantum fluctuations decrease with increasing $S$,
the paramagnetic phase must eventually disappear for large spin
and give way to a first order transition from N\'eel to the stripe state at $ J_2 = 0.5 J_1 $
as in the classical system at $ S \to \infty $.
In this work,
we use the spin FRG to compute the ground state phase diagram of the $ J_1 $-$ J_2 $ model for arbitrary spin $S$
and thereby show that the paramagnetic phase disappears for spin $ S \gtrsim 5 $.

The rest of this paper is organized as follows:
In Sec.~\ref{sec:FRG},
we introduce the basic idea of spin FRG and present the relevant flow equations.
We also discuss how the spin algebra entails both an infinite tower of sum rules for the spin vertices
and nontrivial dynamics, both of which are crucial to develop a 
nonperturbative truncation scheme of the FRG flow equations which allows us to 
detect possible quantum phase transitions.
The resulting truncated system of flow equations is then solved 
numerically in Sec.~\ref{sec:phase_diagram} to obtain the phase diagram of the $J_1$-$J_2$ model.
In the final Sec.~\ref{sec:outlook} we summarize our main results and discuss 
how it could be extended in future work. 
Additional technical details are presented  in three Appendices:
In Appendix \ref{app:temperature} we explain in detail how the spin FRG 
flow equations are rescaled such that the flow directly generates the temperature dependence of spin correlation functions.
Appendix \ref{app:sum_rules}
shows how the operator identity $\bm{S}_i^2 = S (S+1)$ implies an infinite tower of sum rules for the
spin vertices which can be elegantly derived from the spin FRG flow equations.
Finally, in Appendix \ref{app:high-frequency} we derive the   
high-frequency approximation of the truncated flow equations.

\section{Non-perturbative  truncation of FRG flow equations}

\label{sec:FRG}

\subsection{Vertex expansion and exact flow equations}

To implement the spin FRG,
we continuously deform the exchange interaction
$ J ( \bm{k} ) \to J_\Lambda ( \bm{k} ) $
with a scale $ \Lambda \in [ 0 , 1 ] $,
such that
$ J_{ \Lambda = 0 } ( \bm{k} ) = 0 $ and
$ J_{ \Lambda = 1 } ( \bm{k} ) = J ( \bm{k} ) $.
At the initial scale $ \Lambda = 0 $,
the system then decouples into an ensemble of isolated quantum spins,
for which one can compute arbitrary correlation functions 
analytically \cite{Vaks1968, Izyumov1988, Goll2019, Halbinger2023}.
The evolution from this initial condition to the full interacting Heisenberg model \eqref{eq:J1-J2}
with increasing $\Lambda$ is then described by the spin FRG flow equations. 
In this work,
we use a simple multiplicative deformation scheme that linearly switches on the 
exchange interaction~\cite{Krieg2019,Tarasevych2022b},
\begin{equation} \label{eq:deformation_scheme}
J_\Lambda ( \bm{k} ) = \Lambda J ( \bm{k} ) \; .
\end{equation}
In the absence of external magnetic fields,
this deformation scheme merely multiplies the Heisenberg model \eqref{eq:J1-J2} by 
the deformation parameter $\Lambda$. 
Thus,
it can also be interpreted as changing the temperature $ T $ to $ T / \Lambda $.
In effect,
the $ \Lambda $-flow is therefore equivalent to a temperature flow \cite{Honerkamp2001}
that starts at $ T \to \infty $ ($ \Lambda = 0 $) and ends at the physical temperature 
$T$ ($ \Lambda = 1 $).
This allows one to extract the full temperature dependence of the spin correlation functions
from a single solution of the FRG flow equations, as explained in more detail in
Appendix \ref{app:temperature}.
In this way,
one can see that the spin FRG flow equations are perturbatively controlled in 
$ J ( \bm{k} ) / T $.
An iterative solution of the flow equations consequently generates 
the high-temperature expansion.
The goal of the spin FRG formalism can thus be understood as an attempt
to extrapolate from the perturbatively controlled high-temperature regime
to low temperatures by re-summing certain classes of diagrams.

Because at the initial scale $ \Lambda = 0 $ each individual spin is conserved,
special care has to be taken in defining suitable vertex functions 
for the dynamic part of the spin fluctuations \cite{Krieg2019, Goll2019, Tarasevych2021}.
Here,
we employ the ``hybrid vertices'' introduced in Ref.~[\onlinecite{Tarasevych2021}]
that are one-line irreducible only in the static sector,
and interaction-irreducible in the dynamic sector.
Their generating functional $ \Gamma_\Lambda [ \bm{m} , \bm{\eta} ] $
then depends both on the static (classical) magnetization $ \bm{m} $ as well as
on a dynamic (quantum) auxiliary field $ \bm{\eta} $.
The latter can be interpreted as the dynamic part of a fluctuating exchange field.
In the paramagnetic phase
of our  spin-rotation invariant Heisenberg model defined in Eq.~\eqref{eq:J1-J2},
this generating functional has the vertex expansion
%
%
%
%
\begin{widetext}
\begin{align}
\Gamma_\Lambda [ \bm{m} , \bm{\eta} ] 
= {} &
N f_\Lambda / T
+ \frac{ 1 }{ 2 T } \int_{ \bm{k} } \left[ 
J ( \bm{k} ) + \Sigma_\Lambda ( \bm{k} )
\right] \bm{m}_{ \bm{k} } \cdot \bm{m}_{ - \bm{k} }
- \frac{ 1 }{ 2 } \int_K \left[ 
G ( \bm{k} ) + \Pi_\Lambda ( K )
\right] \bm{\eta}_{ K } \cdot \bm{\eta}_{ - K }
\nonumber\\[.1cm]
& 
+ \frac{ 1 }{ 2 }
\int_{ \bm{k}_1 } \int_{ K_2 } \int_{ K_3 } \delta( \bm{k}_1 + K_2 + K_3  )
\Gamma_\Lambda^{ x \tilde{y} \tilde{z} } ( \bm{k}_1 , K_2 , K_3 )
\bm{m}_{ \bm{k}_1 } \cdot \left( \bm{\eta}_{ K_2 } \times \bm{\eta}_{ K_3 } \right)
\nonumber\\[.1cm]
& 
+ \frac{ 1 }{ 3! }
\int_{ K_1 } \int_{ K_2 } \int_{ K_3 } \delta( K_1 + K_2 + K_3  )
\Gamma_\Lambda^{ \tilde{x} \tilde{y} \tilde{z} } ( K_1 , K_2 , K_3 )
\bm{\eta}_{ K_1 } \cdot \left( \bm{\eta}_{ K_2 } \times \bm{\eta}_{ K_3 } \right)
\nonumber\\[.1cm]
& 
+ \frac{ 1 }{ 4! T }
\int_{ \bm{k}_1 } \int_{ \bm{k}_2 } \int_{ \bm{k}_3 } \int_{ \bm{k}_4 } 
\delta( \bm{k}_1 + \bm{k}_2 + \bm{k}_3 + \bm{k}_4  )
\sum_{ \alpha_1 \ldots \alpha_4 }
\Gamma_\Lambda^{ \alpha_1 \alpha_2 \alpha_3 \alpha_4 } ( \bm{k}_1 , \bm{k}_2 , \bm{k}_3 , \bm{k}_4 )
m_{ \bm{k}_1 }^{ \alpha_1 } m_{ \bm{k}_2 }^{ \alpha_2 } m_{ \bm{k}_3 }^{ \alpha_3 } m_{ \bm{k}_4 }^{ \alpha_4 }
\nonumber\\[.1cm]
& 
+ \frac{ 1 }{ ( 2! )^2 }
\int_{ \bm{k}_1 } \int_{ \bm{k}_2 } \int_{ K_3 } \int_{ K_4 } \delta( \bm{k}_1 + \bm{k}_2 + K_3 + K_4  )
\sum_{ \alpha_1 \ldots \alpha_4 }
\Gamma_\Lambda^{ \alpha_1 \alpha_2 \tilde{\alpha}_3 \tilde{\alpha}_4 } ( \bm{k}_1 , \bm{k}_2 , K_3 , K_4 )
m_{ \bm{k}_1 }^{ \alpha_1 } m_{ \bm{k}_2 }^{ \alpha_2 } \eta_{ K_3 }^{ \alpha_3 } \eta_{ K_4 }^{ \alpha_4 }
\nonumber\\[.1cm]
& 
+ \frac{ 1 }{ 3! }
\int_{ \bm{k}_1 } \int_{ K_2 } \int_{ K_3 } \int_{ K_4 } \delta( \bm{k}_1 + K_2 + K_3 + K_4  )
\sum_{ \alpha_1 \ldots \alpha_4 }
\Gamma_\Lambda^{ \alpha_1 \tilde{\alpha}_2 \tilde{\alpha}_3 \tilde{\alpha}_4 } ( \bm{k}_1 , K_2 , K_3 , K_4 )
m_{ \bm{k}_1 }^{ \alpha_1 } \eta_{ K_2 }^{ \alpha_2 } \eta_{ K_3 }^{ \alpha_3 } \eta_{ K_4 }^{ \alpha_4 }
\nonumber\\[.1cm]
&
+ \frac{ 1 }{ 4! }
\int_{ K_1 } \int_{ K_2 } \int_{ K_3 } \int_{ K_4 } \delta( K_1 + K_2 + K_3 + K_4  )
\sum_{ \alpha_1 \ldots \alpha_4 }
\Gamma_\Lambda^{ \tilde{\alpha}_1 \tilde{\alpha}_2 \tilde{\alpha}_3 \tilde{\alpha}_4 } ( K_1 , K_2 , K_3 , K_4 )
\eta_{ K_1 }^{ \alpha_1 } \eta_{ K_2 }^{ \alpha_2 } \eta_{ K_3 }^{ \alpha_3 } \eta_{ K_4 }^{ \alpha_4 }
+ \ldots \; .
\label{eq:vertex_expansion}
\end{align}
\end{widetext}
Here,
$ f_\Lambda $ is the scale-dependent free energy density,
the dots stand for terms with more than four fields,
and $ K = ( \bm{k} , \omega ) $ collects wave vectors $ \bm{k} $ and Matsubara frequencies $ \omega $.
The associated integration and delta symbols are
$ \int_K = T \sum_\omega \int_{ \bm{k} } = ( T / N ) \sum_{ \bm{k} \omega } $ and
$ \delta( K ) = ( 1 / T ) \delta_{ \omega , 0 } \delta( \bm{k} ) = ( N / T ) \delta_{ \omega, 0 } \delta_{ \bm{k} , 0 } $,
respectively.
We also use the shorthand notation $ \bm{k}_1 + K_2 = ( \bm{k}_1 + \bm{k}_2 , \omega_2 ) $.
Greek superscripts denote Cartesian spin components $ \alpha , \tilde{\alpha} \in \{ x , y , z \} $,
where $ \tilde{\alpha} $ denotes dynamic $ \eta^\alpha_K $ external legs,
whereas $ \alpha $ refers to static $ m^\alpha_{ \bm{k} } $ legs.
Note also that spin-rotation invariance demands that only 4-point vertices with pairwise identical spin component labels are finite,
that is
\begin{align}
&
\Gamma_\Lambda^{ \alpha_1 \alpha_2 \alpha_3 \alpha_4 } ( \bm{k}_1 , \bm{k}_2 , \bm{k}_3 , \bm{k}_4 ) 
\nonumber\\[.1cm]
= {} &
\delta^{ \alpha_1 \alpha_2 } \delta^{ \alpha_3 \alpha_4 }
\Gamma_\Lambda^{ x x y y } ( \bm{k}_1 , \bm{k}_2 , \bm{k}_3 , \bm{k}_4 )
\nonumber\\
&
+
\delta^{ \alpha_1 \alpha_3 } \delta^{ \alpha_2 \alpha_4 }
\Gamma_\Lambda^{ x x y y } ( \bm{k}_1 , \bm{k}_3 , \bm{k}_2 , \bm{k}_4 )
\nonumber\\
&
+
\delta^{ \alpha_1 \alpha_4 } \delta^{ \alpha_2 \alpha_3 }
\Gamma_\Lambda^{ x x y y } ( \bm{k}_1 , \bm{k}_4 , \bm{k}_2 , \bm{k}_3 ) \; ,
\end{align}
and likewise for the mixed classical-quantum and quantum 4-point vertices.

The scale-dependent static spin self-energy $ \Sigma_\Lambda ( \bm{k} ) $ determines 
the scale-dependent static spin susceptibility via
\begin{equation} \label{eq:g_static}
G_\Lambda ( \bm{k} ) = \frac{ 1 }{ J_\Lambda ( \bm{k} ) + \Sigma_\Lambda ( \bm{k} ) } \; .
\end{equation}
At the initial scale $\Lambda =0$ corresponding to isolated spins, 
the spin self-energy is
\begin{equation} \label{eq:sigma_initial}
\Sigma_0 = \frac{ 3 T }{ S ( S + 1 ) } \; ,
\end{equation}
which can be identified with the inverse susceptibility of a free spin.
The dynamic spin susceptibility $ G_\Lambda ( \bm{k} , \omega \neq 0 ) $ \cite{footnoteG}
is on the other hand parametrized via the irreducible dynamic spin susceptibility 
$ \Pi_\Lambda ( K ) $ \cite{Tarasevych2021} as
\begin{equation} \label{eq:g_dynamic}
G_\Lambda ( K ) = \frac{ \Pi_\Lambda ( K ) }{ 1 + G_\Lambda^{ - 1 } ( \bm{k} ) \Pi_\Lambda ( K ) } \; .
\end{equation}
For the isolated spins at $ \Lambda = 0 $,
spin conservation implies the initial condition $ \Pi_0 ( \omega \neq 0 ) = 0 $.

\begin{figure}
\includegraphics[width=\linewidth]{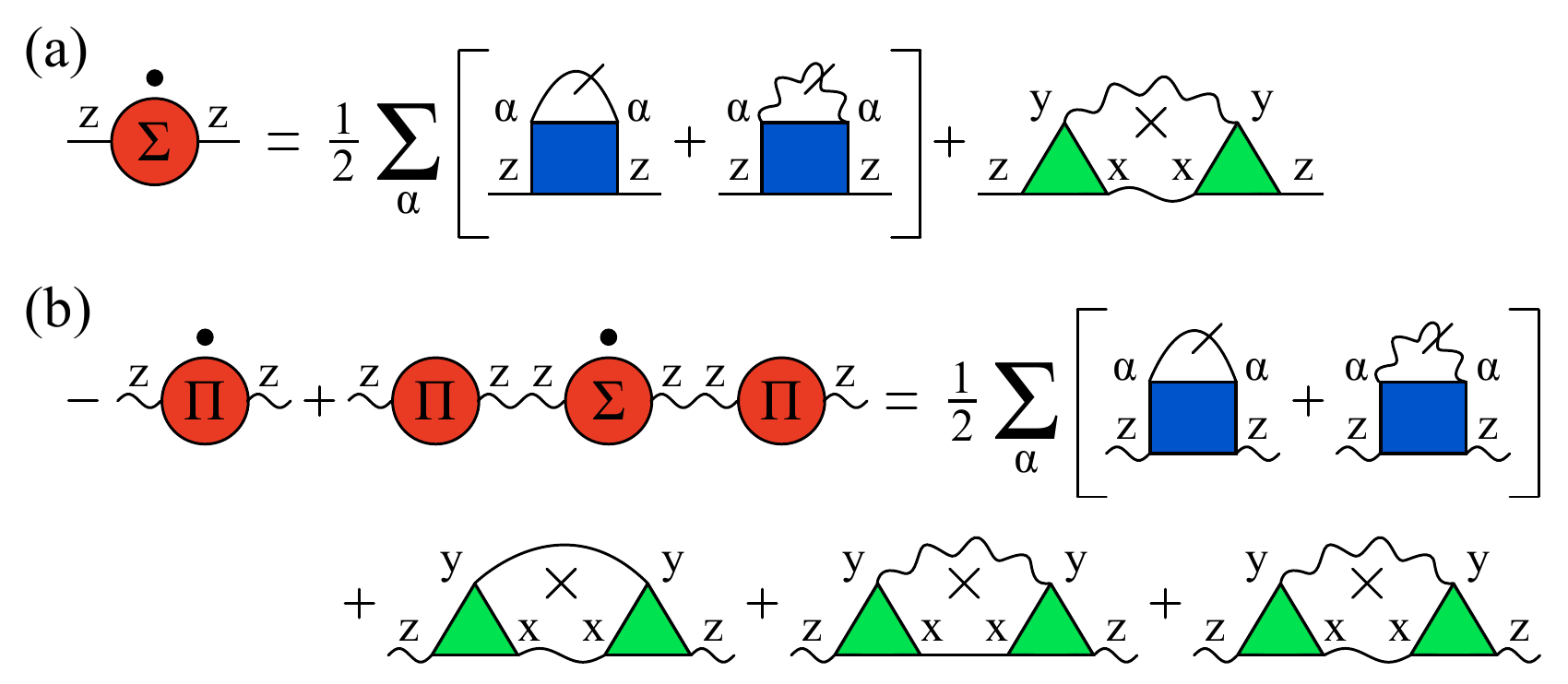}
\caption{Graphical representations of the flow equations
for (a) the static spin self-energy, Eq.~\eqref{eq:sigma_flow}, 
and (b) the dynamic spin susceptibility, Eq.~\eqref{eq:pi_flow}.
Colored symbols represent $n$-point vertices,
with static and dynamic external legs denoted by straight and wavy lines, respectively.
The corresponding internal lines represent the static spin susceptibility $ G_\Lambda ( \bm{k} ) $
and the dynamic propagator $ F_\Lambda ( K ) $.
Slashed lines denote the associated single-scale propagators \cite{Kopietz2010, Tarasevych2021}.
Crosses inside loops mean that every internal line is successively replaced by its single-scale counterpart.
Dots above vertices stand for scale derivatives $ \partial_\Lambda $.}
\label{fig1}
\end{figure}
The spin FRG flow equations for the spin self-energy and polarization 
defined via Eqs.~\eqref{eq:g_static} and \eqref{eq:g_dynamic}, respectively,
are explicitly given by \cite{Tarasevych2021}
\begin{align} 
\partial_\Lambda \Sigma_\Lambda ( \bm{k} ) 
= {} &
- T \int_{ \bm{q} } \left[ \partial_\Lambda J_\Lambda ( \bm{q} ) \right] G_\Lambda^2 ( \bm{q} )
\nonumber\\
&   \hspace{-5mm} \times
\left[
\frac{ 1 }{ 2 } \sum_\alpha \Gamma_\Lambda^{ z z \alpha \alpha } ( - \bm{k} , \bm{k} , \bm{q} , - \bm{q} )
+ \gamma_\Lambda ( \bm{k} , \bm{q} )
\right] ,
\label{eq:sigma_flow}
\end{align}
and
\begin{align}
&
- \partial_\Lambda \Pi_\Lambda ( K ) 
+ \Pi_\Lambda^2 ( K )  \partial_\Lambda \Sigma_\Lambda ( \bm{k} )
\nonumber\\
= {} &
- T \int_{ \bm{q} } \left[ \partial_\Lambda J_\Lambda ( \bm{q} ) \right] G_\Lambda^2 ( \bm{q} )
\tilde{\gamma}_\Lambda ( K , \bm{q} ) \; ,
\label{eq:pi_flow}
\end{align}
where the contributions of the dynamic vertices are 
\begin{widetext}
\begin{equation} \label{eq:sigma_flow_dynamic}
\gamma_\Lambda ( \bm{k} , \bm{q} )
=  
\sum_{ \nu \neq 0 } F_\Lambda^2 ( Q ) \Biggl\{
\frac{ 1 }{ 2 } \sum_\alpha 
\Gamma_\Lambda^{ z z \tilde{\alpha} \tilde{\alpha} } ( - \bm{k} , \bm{k} , Q , - Q )
- 2 F_\Lambda ( Q + \bm{k} )
\left[ \Gamma_\Lambda^{ x \tilde{y} \tilde{z} } ( \bm{k} , - Q - \bm{k} , Q ) \right]^2
\Biggr\} 
\end{equation}
and
\begin{align}
\tilde{\gamma}_\Lambda ( K , \bm{q} )
= {} & 
\frac{ 1 }{ 2 } 
\sum_\alpha  \biggl\{ \Gamma_\Lambda^{ \alpha \alpha \tilde{z} \tilde{z} } ( - \bm{q} , \bm{q} , K , - K )
+ \sum_{ \nu \neq 0 } F_\Lambda^2 ( Q ) 
\Gamma_\Lambda^{ \tilde{\alpha} \tilde{\alpha} \tilde{z} \tilde{z} } ( - Q , Q , K , - K )
\biggr\}
- 2 F_\Lambda ( \bm{q} + K ) \left[ \Gamma_\Lambda^{ x \tilde{y} \tilde{z} } ( \bm{q} , - \bm{q} - K , K ) \right]^2
\nonumber\\[.1cm]
&
+ 2 F_\Lambda^2 ( \bm{q} , \omega ) G_\Lambda ( \bm{q} + \bm{k} ) 
\left[ \Gamma_\Lambda^{ x \tilde{y} \tilde{z} } ( - \bm{q} - \bm{k} , \bm{q} + \omega , K ) \right]^2
- 2 \sum_{ \nu \neq 0 , - \omega } F_\Lambda^2 ( Q )
F_\Lambda ( Q + K ) \left[ \Gamma_\Lambda^{ \tilde{x} \tilde{y} \tilde{z} } ( - Q - K , Q , K ) \right]^2
\; .
\label{eq:pi_flow_dynamic}
\end{align}
\end{widetext}
Here,
$ Q = ( \bm{q} , \nu ) $, 
$ \bm{q} + K = ( \bm{q} + \bm{k} , \omega ) $,
$ \bm{q} + \omega = ( \bm{q} , \omega ) $, etc.,
and the dynamic propagator is
\begin{equation}
F_\Lambda ( K ) = \frac{ 1 }{ G_\Lambda ( \bm{k} ) + \Pi_\Lambda ( K ) } \; .
\end{equation}
Graphical representations of the flow equations \eqref{eq:sigma_flow} and \eqref{eq:pi_flow} 
are displayed in Fig.~\ref{fig1}.
Note that these flow equations for the 2-point vertices are not closed:
they depend on 3- and 4-point vertices,
which in turn depend on 5- and 6-point vertices,
and so on.
In order to develop a nonperturbative truncation of this infinite hierarchy
that is capable to describe also the low-temperature regime,
we next explore the intricate relationships between the vertices 
implied by the spin algebra.

\subsection{Spin-length sum rules}

\label{sec:sum_rules}

For each lattice site $i$ the spin-operators satisfy the constraint
\begin{equation} \label{eq:length_constraint}
\bm{S}_i^2 = S ( S + 1 ) \; .
\end{equation}
In spin-wave theory this constraint gives rise to the so-called
kinematic interactions between spin waves. In the magnetically ordered phase of three-dimensional ferromagnets the effect of these interactions has been thoroughly
investigated by  Dyson \cite{Dyson1956}, who concluded 
that  kinematic  interactions do not contribute to the
low-temperature thermodynamics. 
On the other hand, in reduced dimensions or for frustrated quantum magnets
the role of the spin-length constraint is expected to be more important.
In fact, the low-temperature properties of quantum antiferromagnets in two dimensions
have been successively modeled
by a nonlinear $\sigma$-model \cite{Chakravarty1988, Chakravarty1989, Hasenfratz1991, Hasenfratz1993, Chubukov1994} where 
interactions between spin fluctuations  arise 
solely from the spin-length constraint.
This suggests that a sensible  truncation of the spin FRG flow equations
for the two-dimensional $J_1$-$J_2$ quantum Heisenberg model  
should also incorporate this constraint.
For the scale-dependent spin susceptibility,
this means that
\begin{equation} \label{eq:g_sum_rule}
\int_K G_\Lambda ( K ) = \frac{ S ( S + 1 ) }{ 3 } = \frac{ T }{ \Sigma_0 } \; .
\end{equation}
Moreover, one can also use the spin-length constraint \eqref{eq:length_constraint} 
to relate higher-order $n$-point to $(n-2)$-point spin correlation functions, which
actually entails an infinite tower of sum rules for the vertices.
We show in Appendix \ref{app:sum_rules} that these sum rules can be generated
efficiently from the exact spin FRG flow equations 
for the vertices in the expansion (\ref{eq:vertex_expansion})
by the following replacements: 
\begin{enumerate}[label=(\roman*)]
\item
$ \partial_\Lambda J_\Lambda ( \bm{q} ) \to 1 $,
\item
$ \partial_\Lambda \Sigma_\Lambda ( \bm{k} ) \to - 1 $, 
\item
$ \partial_\Lambda f_\Lambda \to S ( S + 1 ) / 2 $,
\item
all other $\Lambda$-derivatives are set to zero.
\end{enumerate}
Applied to the flow equations \eqref{eq:sigma_flow} and \eqref{eq:pi_flow} for the
$2$-point vertices
this yields the two additional sum rules 
\begin{subequations}
\begin{align}
&
1 
=
T \int_{ \bm{q} } G_\Lambda^2 ( \bm{q} )
\left[
\frac{ 1 }{ 2 } \sum_\alpha \Gamma_\Lambda^{ z z \alpha \alpha } ( - \bm{k} , \bm{k} , \bm{q} , - \bm{q} )
+ \gamma_\Lambda ( \bm{k} , \bm{q} )
\right] 
\; ,
\label{eq:sigma_sum_rule}
\\[.1cm]
&
\Pi_\Lambda^2 ( K ) 
= 
T \int_{ \bm{q} } G_\Lambda^2 ( \bm{q} )
\tilde{\gamma}_\Lambda ( K , \bm{q} ) \; .
\label{eq:pi_sum_rule}
\end{align}
\end{subequations}
At the initial scale $ \Lambda = 0 $,
these two sum rules reduce to the initial values 
\begin{equation} \label{eq:4-point_initial}
\Gamma_0^{ z z z z } = 3 \Gamma_0^{ x x y y } = 
\frac{ \Sigma_0^2 }{ 5 T^2 } \left( 6 T + \Sigma_0 \right)
\end{equation}
and $ \Pi_0 ( \omega ) = 0 $.
Note that Eq.~\eqref{eq:4-point_initial} expresses the exact static 4-point vertex
of an isolated spin in terms of its inverse susceptibility $ \Sigma_0$ given in Eq.~\eqref{eq:sigma_initial}
as a consequence of the finite length of the spin operator.
By considering the flow of the susceptibility sum rule \eqref{eq:g_sum_rule},
one can also show that it is automatically satisfied as long as the two
sum rules \eqref{eq:sigma_sum_rule} and \eqref{eq:pi_sum_rule} are fulfilled exactly.
However,
as soon as one truncates the flow equations and applies the inevitable approximations,
this no longer holds in general.
This enables us to take the three sum rules as independent constraints on the vertices.

The dynamic sum rule, Eq.~\eqref{eq:pi_sum_rule}, 
can be plugged back into the associated flow equation \eqref{eq:pi_flow},
which becomes
\begin{equation} \label{eq:pi_flow1}
\partial_\Lambda \Pi_\Lambda ( K ) 
= 
T \int_{ \bm{q} } \left[ 
\partial_\Lambda J_\Lambda ( \bm{q} ) + \partial_\Lambda \Sigma_\Lambda ( \bm{k} ) \right] G_\Lambda^2 ( \bm{q} )
\tilde{\gamma}_\Lambda ( K , \bm{q} ) \; .
\end{equation}
To implement also the static sum rule, Eq.~\eqref{eq:sigma_sum_rule}, 
we need a suitable approximation for the static 4-point vertex.
Here,
we are guided by the insight that for non-degenerate ground states,
it is sufficient to neglect the momentum dependence of the static 4-point vertex \cite{Tarasevych2022b}.
On the other hand,
we expect that in a frustrated system like the $J_1$-$J_2$ model,
the momentum-dependent quantum fluctuations encoded in $ \gamma_\Lambda ( \bm{k} , \bm{q} ) $
are crucial.
The static sum rule \eqref{eq:sigma_sum_rule} now tells us that in the presence of such quantum fluctuations, 
a momentum-independent static 4-point vertex $ U_\Lambda $ is no longer sufficient 
to satisfy the spin-length constraint \eqref{eq:length_constraint}.
Instead, the static 4-point vertex must be renormalized 
by the momentum-dependent quantum fluctuations $ \gamma_\Lambda ( \bm{k} , \bm{q} ) $.
In order to implement this renormalization,
we make the following ansatz for the static 4-point vertex,
%
%
\begin{equation}
\frac{ 1 }{ 2 } \sum_\alpha \Gamma_\Lambda^{ z z \alpha \alpha } ( - \bm{k} , \bm{k} , \bm{q} , - \bm{q} ) =
U_\Lambda
+ \left[ V_\Lambda ( \bm{k} ) - 1 \right] \gamma_\Lambda ( \bm{k} , \bm{q} ) \; ,
\end{equation}
with initial conditions 
$ U_0 = 5 \Gamma_0^{ z z z z } / 6 $ and $ V_0 ( \bm{k} ) = 1 $.
This turns the flow equation \eqref{eq:sigma_flow} into
\begin{align} 
\partial_\Lambda \Sigma_\Lambda ( \bm{k} ) 
= {} &
- T \int_{ \bm{q} } \left[ \partial_\Lambda J_\Lambda ( \bm{q} ) \right] G_\Lambda^2 ( \bm{q} )
\nonumber\\
& \times
\left[
U_\Lambda + V_\Lambda ( \bm{k} ) \gamma_\Lambda ( \bm{k} , \bm{q} )
\right] 
\; .
\label{eq:sigma_flow1}
\end{align}
Note that the new coupling $V_\Lambda ( \bm{k} )$ controls the strength 
of the quantum fluctuations.
It is fixed by demanding that the static sum rule \eqref{eq:sigma_sum_rule} holds,
yielding
\begin{equation} \label{eq:v_sum_rule}
V_\Lambda ( \bm{k} ) = 
\frac{
1 - T U_\Lambda \int_{ \bm{q} } G_\Lambda^2 ( \bm{q} )
}{
T \int_{ \bm{q} } G_\Lambda^2 ( \bm{q} ) \gamma_\Lambda ( \bm{k} , \bm{q} )
}
\; .
\end{equation}
The scale-dependence of the remaining coupling $ U_\Lambda $ is finally set by demanding 
that the susceptibility sum rule \eqref{eq:g_sum_rule} is conserved during the flow,
\begin{equation}
\partial_\Lambda \int_K G_\Lambda ( K ) = 0 \; .
\end{equation}
Inserting the flow equations \eqref{eq:pi_flow1} and \eqref{eq:sigma_flow1}
for the 2-point vertices and solving for $ U_\Lambda $ then yields \cite{footnoteUV}

\begin{widetext}
\begin{equation} \label{eq:u_sum_rule}
U_\Lambda =
-
\frac{
\int_K G_\Lambda^2 ( K ) \left[
\partial_\Lambda J_\Lambda ( \bm{k} ) + \dot{\Sigma}_\Lambda^{ ( 1 ) } ( \bm{k} ) \right]
- \int_{ K ( \omega \neq 0 ) } G_\Lambda^2 ( \bm{k} ) F_\Lambda^2 ( K ) \dot{\Pi}_\Lambda^{ ( 1 ) } ( K )
}{
\int_K G_\Lambda^2 ( K ) \dot{\Sigma}_\Lambda^{ ( 0 ) } ( \bm{k} )
- \int_{ K ( \omega \neq 0 ) } G_\Lambda^2 ( \bm{k} ) F_\Lambda^2 ( K ) \dot{\Pi}_\Lambda^{ ( 0 ) } ( K )
} \; ,
\end{equation}
\end{widetext}
where we have parametrized  the flow of the static self-energy as follows,
\begin{subequations} \label{eq:sigma_flow2}
\begin{align}
\partial_\Lambda \Sigma_\Lambda ( \bm{k} ) 
= {} &
U_\Lambda \dot{\Sigma}_\Lambda^{ ( 0 ) } ( \bm{k} ) + \dot{\Sigma}_\Lambda^{ ( 1 ) } ( \bm{k} ) 
\; ,
\\[.1cm]
\dot{\Sigma}_\Lambda^{ ( 0 ) } ( \bm{k} )
= {} &
- T \int_{ \bm{q} } \left[ 
\partial_\Lambda J_\Lambda ( \bm{q} ) + \dot{\Sigma}_\Lambda^{ ( 1 ) } ( \bm{k} )
\right] G_\Lambda^2 ( \bm{q} ) \; ,
\\[.1cm]
\dot{\Sigma}_\Lambda^{ ( 1 ) } ( \bm{k} )
= {} &
-
\frac{ 
\int_{ \bm{q} } \left[ \partial_\Lambda J_\Lambda ( \bm{q} ) \right] G_\Lambda^2 ( \bm{q} )
\gamma_\Lambda ( \bm{k} , \bm{q} )
}{ \int_{ \bm{q} } G_\Lambda^2 ( \bm{q} ) \gamma_\Lambda ( \bm{k} , \bm{q} )  } \; ,
\end{align}
\end{subequations}
and similarly for the flow of the dynamic two-point function,
\begin{subequations} \label{eq:pi_flow2}
\begin{align}
\partial_\Lambda \Pi_\Lambda ( K ) 
= {} &
U_\Lambda \dot{\Pi}_\Lambda^{ ( 0 ) } ( K ) + \dot{\Pi}_\Lambda^{ ( 1 ) } ( K ) 
\; ,
\\[.1cm]
\dot{\Pi}_\Lambda^{ ( 0 ) } ( K )
= {} &
T \dot{\Sigma}_\Lambda^{ ( 0 ) } ( \bm{k} ) \int_{ \bm{q} } G_\Lambda^2 ( \bm{q} )
\tilde{\gamma}_\Lambda ( K , \bm{q} )
\; ,
\\[.1cm]
\dot{\Pi}_\Lambda^{ ( 1 ) } ( K )
= {} &
T \int_{ \bm{q} } \left[ 
\partial_\Lambda J_\Lambda ( \bm{q} ) + \dot{\Sigma}_\Lambda^{ ( 1 ) } ( \bm{k} ) \right] G_\Lambda^2 ( \bm{q} )
\tilde{\gamma}_\Lambda ( K , \bm{q} )
\; .
\end{align}
\end{subequations}
It is apparent from these expressions that satisfying the spin-length constraint \eqref{eq:length_constraint}
entails a highly nonlinear feedback mechanism between the quantum dynamics and the classical statics,
which goes beyond the loop resummations of the flow equations on their own.

\subsection{Dynamic vertices and spin conservation}

\label{sec:dynamics}

Besides the spin-length constraint,
the spin algebra also generates highly nontrivial quantum dynamics.
This is already apparent in the complicated frequency dependence of the
correlation functions and vertices of isolated spins \cite{Goll2019,Tarasevych2021}.
In a truncation of the flow equations,
these also need to be taken into account nonperturbatively
if one wants to retain conservation laws of the Heisenberg model \eqref{eq:J1-J2} throughout the flow, 
such as the spin conservation
$ G_\Lambda ( \bm{k} = 0 , \omega \neq 0 ) = 0 $ \cite{Tarasevych2021},
which implies $ \Pi_\Lambda ( \bm{k} = 0 , \omega \neq 0 ) = 0 $;
see Eq.~\eqref{eq:g_dynamic}.
To develop such a truncation of the dynamic vertices,
we employ the exact equations of motion of the spin correlation functions \cite{Goll2019,Tarasevych2021}.
The equations of motion for the relevant connected 3- and 4-point spin correlation functions are explicitly given by
\begin{subequations}
\begin{align}
&
\omega G_\Lambda^{ x y z } ( - Q - K , Q , K ) 
= 
G_\Lambda ( Q + K ) - G_\Lambda ( Q ) 
\nonumber\\[.1cm]
& \hspace*{1.5cm}
+ \left[ J_\Lambda ( \bm{q} + \bm{k} ) - J_\Lambda ( \bm{q} ) \right] G_\Lambda ( Q + K ) G_\Lambda ( Q ) 
\nonumber\\
& \hspace*{1.5cm}
+ \int_{ Q' } \left[ J_\Lambda ( \bm{q}' + \bm{k} ) - J_\Lambda ( \bm{q}' ) \right]
\nonumber\\
& \hspace*{2cm} \times
G_\Lambda^{ x x y y } ( - Q - K , Q' + K , Q , - Q' ) 
\; , \\[.2cm]
&
\omega G_\Lambda^{ x x y y } ( Q , - Q , - K , K )
= - G_\Lambda^{ x y z } ( Q + K , - Q , - K ) 
\nonumber\\[.1cm]
& \hspace*{.7cm} \times
\left\{ 1 + \left[ J_\Lambda ( \bm{q} + \bm{k} ) - J_\Lambda ( \bm{q} ) \right] G_\Lambda ( Q ) \right\}
+ \left( Q \leftrightarrow - Q \right)
\nonumber\\[.1cm]
& \hspace*{.7cm} 
+ \int_{ Q' } \left[ J_\Lambda ( \bm{q}' + \bm{k} ) - J_\Lambda ( \bm{q}' ) \right]
\nonumber\\
& \hspace*{1.2cm} \times
G_\Lambda^{ x y y y z } ( - K , - Q , Q , Q' + K , - Q' )
\; .
\end{align}
\end{subequations}
Exact equations for the dynamic 3- and 4-point vertices 
can now be obtained from the tree expansion \cite{Kopietz2010}.
Note in particular that for $ \bm{k} = 0 $ all terms involving the exchange
interaction vanish, 
including the loop integrations that couple to higher-order correlation functions.
Only the dynamic terms generated by the spin algebra remain in this case.
In order to develop a minimal truncation consistent with the spin algebra 
and especially the spin conservation at $ \bm{k} = 0 $,
we therefore conclude that the loop integrals may be dropped.
Then the dynamic 3- and 4-point spin correlation functions and hence also the associated vertices 
are entirely determined by the 2-point vertices $ \Sigma_\Lambda ( \bm{k} ) $ and $ \Pi_\Lambda ( K ) $.
However,
by neglecting the loop integrations we incur errors at order $ J_\Lambda^3 $ for $ \bm{k} \neq 0 $.
As both $ \Sigma_\Lambda ( \bm{k} ) - \Sigma_0 $ and $ \Pi_\Lambda ( K ) $ are at least of order $ J_\Lambda^2 $,
we should then for consistency only retain terms up to linear order in these quantities,
which becomes exact for $ \bm{k} = 0 $.
This strategy yields the following minimal approximations 
for the nonvanishing dynamic 3- and 4-point vertices:
%
%
\begin{subequations}
\label{eq:gammaapprox}
\begin{align}
\Gamma_\Lambda^{ x \tilde{y} \tilde{z} } ( \bm{q} , - \bm{q} - K , K )
\approx {} & \frac{ 1 }{ \omega } 
\; , \\[.1cm]
\Gamma_\Lambda^{ \tilde{x} \tilde{y} \tilde{z} } ( - Q - K , Q , K )
\approx {} & \frac{ 1 }{ \omega } \left[ \Pi_\Lambda ( Q + K ) - \Pi_\Lambda ( Q ) \right]
\; , \\[.1cm]
\Gamma_\Lambda^{ x x \tilde{y} \tilde{y} } ( - \bm{q} , \bm{q} , K , - K )
\approx {} & \frac{ 1 }{ \omega^2 } \left[ 
\Sigma_\Lambda ( \bm{q} + \bm{k} ) - \Sigma_\Lambda ( \bm{q} )
\right]
\nonumber\\
&
+ \left( \bm{q} \leftrightarrow - \bm{q} \right)
\; , \\[.1cm]
\Gamma_\Lambda^{ \tilde{x} \tilde{x} \tilde{y} \tilde{y} } ( - Q , Q , K , - K )
\approx {} & \frac{ 1 }{ \omega^2 } \Bigl\{ 
\delta_{ \omega + \nu , 0 } \Pi_\Lambda ( \bm{q} , \omega )
\nonumber\\
& \hspace*{-3.5cm}
+ \left( 1 - \delta_{ \omega + \nu , 0 } \right)
\left[ \Pi_\Lambda ( Q + K ) - \Pi_\Lambda ( Q ) \right]
\Bigr\} 
+ \left( Q \leftrightarrow - Q \right)
\; .
\end{align}
\end{subequations}
A major advantage of these expressions is that they are exact in the limit 
$ \bm{k} \to 0 $ and therefore
satisfy spin conservation exactly. 
On the other hand, for $ \bm{k} \neq 0 $
Eqs.~(\ref{eq:gammaapprox})
 are only correct to order $ J_\Lambda^2 $.
The contributions \eqref{eq:sigma_flow_dynamic} and \eqref{eq:pi_flow_dynamic} of the dynamic vertices 
to the flow equations for the $2$-point vertices then reduce 
to \cite{footnoteclassical}
\begin{subequations} \label{eq:dynamic}
\begin{align}
\gamma_\Lambda ( \bm{k} , \bm{q} ) 
= {} &
2 \sum_{ \nu \neq 0 } \frac{ F_\Lambda^2 ( \bm{q} , \nu ) }{ \nu^2 }
\bigl[  
\Sigma_\Lambda ( \bm{k} + \bm{q} ) - \Sigma_\Lambda ( \bm{k} )
\nonumber\\
& \phantom{ aaaaaaaaaaaa }
- F_\Lambda ( \bm{q} + \bm{k} , \nu ) 
\bigr] \; ,
\label{eq:dynamic_a}
\\[.1cm]
\tilde{\gamma}_\Lambda ( K , \bm{q} )
= {} &
\frac{ 2 }{ \omega^2 } 
\Bigl\{
\Sigma_\Lambda ( \bm{q} + \bm{k} ) - \Sigma_\Lambda ( \bm{q} )
\nonumber\\[.15cm]
& \phantom{ aaa }
+  \sum_{ \nu } 
F_\Lambda^2 ( Q ) F_\Lambda ( Q + K )
\nonumber\\
& \phantom{ aaaaa } \times
\left[ G_\Lambda ( \bm{q} + \bm{k} ) + \Pi_\Lambda ( Q ) \right]
\nonumber\\
& \phantom{ aaaaa } \times
\left[ \Pi_\Lambda ( Q + K ) - \Pi_\Lambda ( Q ) \right]
\Bigr\} \; .
\end{align}
\end{subequations}
Spin conservation is ensured by construction because
$ \tilde{\gamma}_\Lambda ( K , \bm{q} ) $ vanishes for $\bm{k} =0$.
Note also that the property $ \tilde{\gamma}_\Lambda ( K , \bm{q} ) \propto 1 / \omega^2 $ implies that
the ergodicity condition $ \Pi_\Lambda^{-1} ( \bm{k} , \omega \to 0 ) \to 0 $ \cite{Tarasevych2021}
is also automatically fulfilled.

This completes our nonperturbative truncation of the FRG flow equations.
The remaining system of flow equations, 
Eqs.~(\ref{eq:u_sum_rule}-\ref{eq:pi_flow2}) together with Eqs.~\eqref{eq:dynamic},
is closed on the level of the 2-point vertices 
$ \Sigma_\Lambda ( \bm{k} ) $ and $ \Pi_\Lambda ( K ) $,
while satisfying both spin-length sum rules and spin conservation.
However,
as we are here mainly interested in the static spin susceptibility
and not in the dynamics on their own,
we additionally apply a high-frequency expansion to the dynamic polarization,
\begin{equation} \label{eq:Pi_via_A}
\Pi_\Lambda ( \bm{k} , \omega ) \approx \frac{ A_\Lambda ( \bm{k} ) }{ \omega^2 } 
 \left[ 1 + {\cal O}\left( \frac{ 1 }{ \omega^2 } \right) \right] \; .
\end{equation}
Anticipating that this approximation will capture the main contribution
to the dynamics over a wide temperature range \cite{Tarasevych2022b},
we can then perform all Matsubara sums analytically.
The resultant high-frequency approximation of the system
of flow equations (\ref{eq:u_sum_rule}-\ref{eq:pi_flow2}) 
is listed explicitly in Appendix \ref{app:high-frequency}.
It turns out that for the $J_1$-$J_2$ model with frustration roughly in the range $ 0.3 \lesssim J_2 / J_1 \lesssim 0.9 $,
the high-frequency approximation \eqref{eq:Pi_via_A}
leads to unphysical results at very low temperatures:
for small $ \bm{k} $ the function
$ A_\Lambda ( \bm{k} ) $ can become negative,
leading to instabilities.
Since the dynamic polarization $ \Pi_\Lambda ( \bm{k} , \omega ) $ has to be positive definite \cite{Tomita1971},
this signals the breakdown of our approximation scheme.
We therefore stop the flow whenever $ A_\Lambda ( \bm{k} ) < 0 $ for some $ \bm{k} $.
For the frustrated model with spin $ S = 1/2 $ for example,
this means that depending on the strength of the frustration, 
we cannot reach temperatures lower than some temperature 
$ T_{\rm min}$ which is typically in the range between $ J_1 / 13 $ and  $ J_1 / 8 $.
However,
we show in the next section that this is sufficient to obtain the full phase diagram of the $J_1$-$J_2$ model.
We have further checked that the high-frequency approximation \eqref{eq:Pi_via_A} is reliable by numerically solving the flow equations without it with up to $50$ positive Matsubara frequencies for several representative values of $ J_2 / J_1 $,
finding no appreciable changes to the results for the static self-energy.

%
%
%
%
%
%
%
%
%
%
%
%
%

\section{Phase diagram}

\label{sec:phase_diagram}

\begin{figure*}
\includegraphics[width=\linewidth]{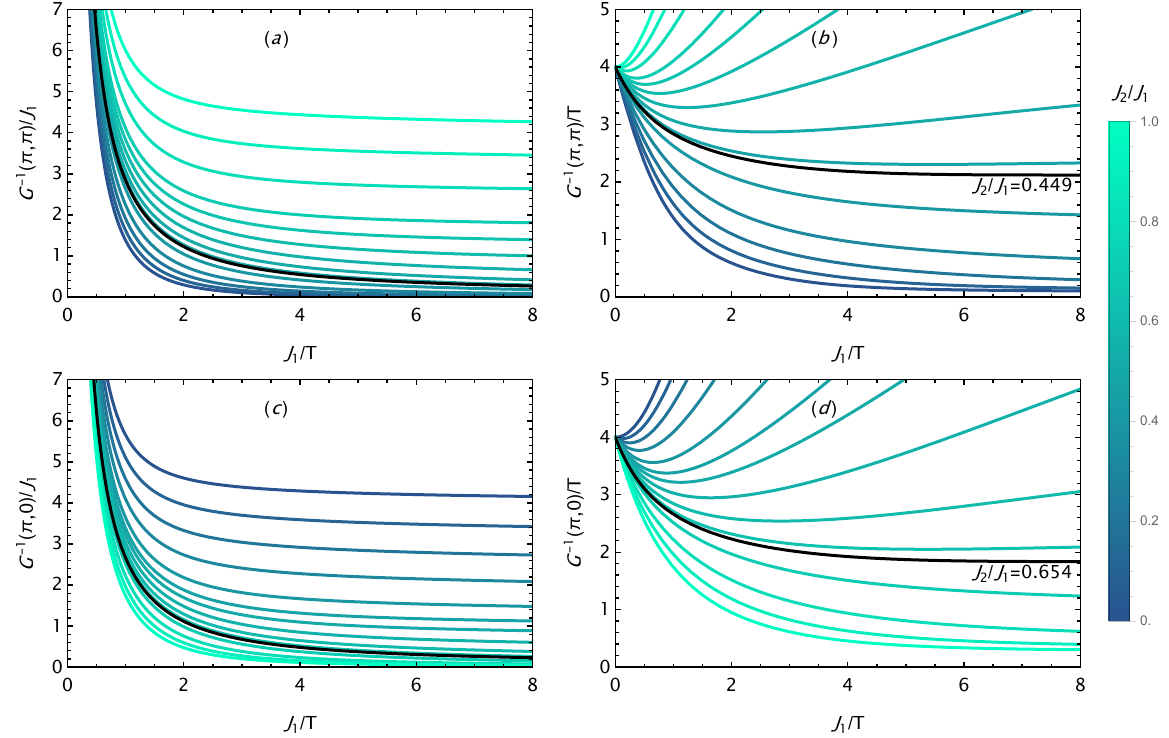}
\caption{Temperature dependence of the inverse spin susceptibility for spin $ S = 1 / 2 $ 
at the N\'eel (a-b) and stripe (c-d) ordering vectors,
$ \bm{Q} = ( \pi , \pi ) $ and $ \bm{Q} = ( \pi , 0 ) $ respectively,
for frustration values $ 0 \le J_2 / J_1 \le 1 $.
(a) and (c) display the absolute value of $ G^{ - 1 } ( \bm{Q} ) $,
whereas (b) and (d) show the value relative to the temperature $T$.
The latter depiction facilitates the distinction between gapless states, where 
$ G^{ - 1 } ( \bm{Q} ) / T $ approaches a constant for large $ J_1 / T $,
and gapped states,
where it increases linearly.
The phase boundaries of Fig.~\ref{fig3} are determined as the values of $ J_2 / J_1 $
for which $ G^{ - 1 } ( \bm{Q} ) / T $ is asymptotically flat at low temperatures.
These special curves are plotted as black solid lines.
We show the temperature dependence only up to $ J_1 / T = 8 $
because at lower temperatures the dynamic coefficient $ A_\Lambda ( \bm{k} ) $ 
flows to unphysical negative values for highly frustrated values of $ J_2 / J_1 $.}
\label{fig2}
\end{figure*}
The solution of the spin FRG flow equations (\ref{eq:u_sum_rule}-\ref{eq:pi_flow2})
directly generates the temperature dependence of physical spin correlation functions,
in particular also of the static spin susceptibility $ G ( \bm{k} ) \equiv G_{ \Lambda = 1 } ( \bm{k} ) $.
Phase transitions from the paramagnetic high-temperature state to a magnetic state
with ordering vector $ \bm{Q} $ can then be identified by the divergence
of $ G ( \bm{Q} ) $ or equivalently the vanishing of the gap $ G^{ - 1 } ( \bm{Q} ) $ at the transition temperature.
We show the temperature dependence of the inverse spin susceptibility at the two classical ordering vectors 
in Fig.~\ref{fig2}
for spin $ S = 1 / 2 $ and frustration values $ 0 \le J_2 / J_1 \le 1 $.
In accordance with the Mermin-Wagner theorem \cite{Mermin1966},
we observe no phase transition at finite temperatures.
In order to identify possible quantum phase transitions at zero temperature,
we have to keep in mind that the spin FRG flow approaches the limit $T \to 0$
only asymptotically.
Therefore we have to analyze the behavior of the gap at the lowest available temperatures.
A convenient way to do this is to plot $ G^{ - 1 } ( \bm{Q} ) / T $ as function of $ J_1 / T $;
see Figs.~\ref{fig2}(b) and (d).
For a gapped state with finite $ G^{ - 1 } ( \bm{Q} ) $ at $ T = 0 $,
this curve grows linearly at large $ J_1 / T $.
If the ground state on the other hand possesses long-range order,
$ G^{ - 1 } ( \bm{Q} ) / T $ approaches a constant.
In this manner,
we find that at small next-nearest neighbor coupling $ J_2 < 0.449 J_1 $,
the gap of the N\'eel state vanishes for $ T \to 0 $,
suggesting the expected N\'eel ordered ground state.
In the opposite limit of large next-nearest neighbor coupling, $ J_2 > 0.654 J_1 $,
we likewise find the expected stripe ordered ground state.
In the intermediate regime   $ 0.449 \le J_2 / J_1 \le 0.654 $
of strong frustration 
the system remains gapped.
We therefore identify the ground state in this regime as a quantum paramagnet
without magnetic long-range order,
in good quantitative agreement with the literature \cite{Oliveira1991, Oitmaa1996, Kotov1999, Capriotti2000, Darradi2008, Isaev2009, Jiang2012, Hu2013, Gong2014, Wang2018, Ferrari2020, Nomura2021}. 
\begin{figure}
\includegraphics[width=\linewidth]{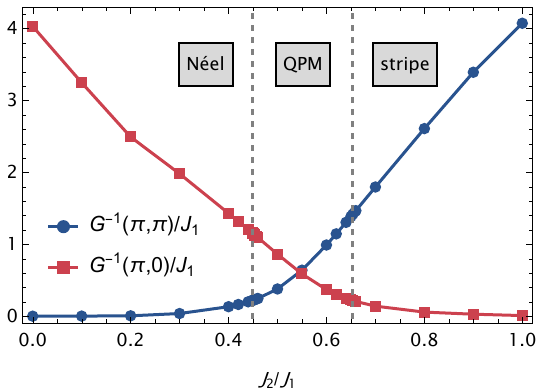}
\caption{Residual gaps of the N\'eel and stripe states, 
$ \bm{Q} = ( \pi , \pi ) $ and $ \bm{Q} = ( \pi , 0 ) $ respectively,
at the end of the spin FRG flow,
for spin $ S = 1 / 2 $.
In the regime of strong frustration for $ 0.449 \le J_2 / J_1 \le 0.654 $,
marked by the dashed gray lines,
we identify a gapped quantum paramagnet (QPM).
Solid lines are guides to the eye.}
\label{fig3}
\end{figure}
The residual gaps of the N\'eel and stripe states at the end of the spin FRG flow are displayed in Fig.~\ref{fig3},
where one clearly sees the emergence of a gapped quantum paramagnet at strong frustration.
Note also that the point where the states with N\'eel and stripe ordering are degenerate
is shifted upwards from the classical value $ J_2 / J_1 = 0.5 $ to $ J_2 / J_1 \approx 0.54 $
by the strong quantum fluctuations.

While our spin FRG flow predicts the existence of the intermediate paramagnetic phase,
it unfortunately does not make any statements about its nature.
Thus, we cannot at this stage distinguish between the possible spin liquid and valence bond states
discussed in the literature \cite{Figueirido1990, Gelfand1989, Oliveira1991, Read1991, Zhitomirksy1996, Jongh2000, Kotov2000, Sushkov2001, Isaev2009, Jiang2012, Hu2013, Gong2014, Wang2018, Nomura2021, Qian2023}.
We also cannot exclude the possibility of a gapless spin liquid state \cite{Hu2013, Wang2018, Ferrari2020} at small $ J_2 $,
since it is hard to distinguish from the N\'eel state in our approach.
This is however not an intrinsic limitation of the spin FRG,
but rather of our truncation,
which focuses on the 2-point spin correlation function.
A more sophisticated truncation capable of investigating dimer correlation functions
would have to include the momentum and frequency dependence of the dynamic 4-point vertex in an unbiased manner,
which is beyond the scope of this work.
An alternative approach would be to add small perturbations to the exchange interaction
that favor certain types of order. 
This was done in Ref.~\cite{Reuther2010} for the pseudofermion flow.
As these perturbations necessarily break translational invariance,
we do not pursue this approach here.
We should also mention that we cannot draw any reliable conclusions about the order of the phase transitions from the gap values in Fig.~\ref{fig3}. 
The reason for this is that we stop the flow at a finite (small) temperature,
which washes out the sharp step expected for a first order transition.

\begin{figure}
\includegraphics[width=\linewidth]{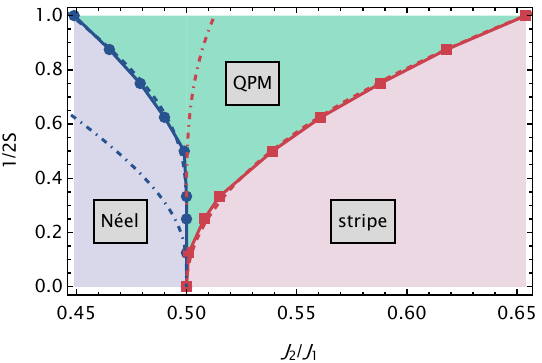}
\caption{Ground state phase diagram of the $J_1$-$J_2$ model as function of frustration $ J_2 / J_1 $
and spin quantum number $1/2S$, 
exhibiting N\'eel, stripe and intermediate quantum paramagnet (QPM) phases.
Solid lines are guides to the eye.
The dashed lines are the fits \eqref{eq:boundaries} for the phase boundaries.
Dash-dotted lines of the corresponding colors represent the linear spin-wave theory prediction of Ref.~[\onlinecite{Chandra1988}] 
for the phase boundaries, which significantly underestimates the extent of the paramagnetic phase for small $S$.}
\label{fig4}
\end{figure}
A major advantage of the spin FRG approach to quantum spin systems 
is that the spin quantum number $S$ only enters as a parameter into the vertex functions.
We can therefore investigate arbitrary values (including 
non-integer and non-half-integer values) 
of $S$ without any additional numerical or analytical effort.
This allows us to tune not only the frustration 
but also the importance of quantum fluctuations,
which vanish for $ S \to \infty $.
The resulting phase diagram as function of frustration $ J_2 / J_1 $
and spin quantum number $1/2S$ is shown in Fig.~\ref{fig4}.
Similar to spin-wave theory \cite{Chandra1988},
our spin FRG predicts the existence of the quantum paramagnetic phase
for spin $S \lesssim 5$.
For even larger spin,
only the two classically ordered phases remain,
with a phase transition at the point of classical degeneracy $ J_2 / J_1 = 0.5 $. 
Qualitatively,
the behavior of the phase boundaries is also as predicted by linear spin wave theory \cite{Chandra1988};
the spin FRG phase boundaries are well approximated by
\begin{subequations} \label{eq:boundaries}
\begin{align}
\frac{1}{2S} & = 
\frac{ - 2.95 
}{ \ln \left( 1/2 - J_2 / J_1 \right) } \; ,
\;\;\; \;\;\;
(\text{N\'eel}) \; ,
\\[.1cm]
\frac{1}{2S} & = 
2.55 \sqrt{ J_2 / J_1 - 1/2 } \; ,
\;\;\; \;\;\;
(\text{stripe}) \; .
\end{align}
\end{subequations}
These curves are shown as dashed blue and red lines in Fig.~\ref{fig4}, respectively.
However, 
the quantitative picture is rather different from the predictions
of spin-wave theory~\cite{Chandra1988}
that are for comparison plotted as dash-dotted lines in Fig.~\ref{fig4},
with a significantly larger and rather asymmetric paramagnetic region that is wider for $ J_2 / J_1 > 0.5 $
than for $ J_2 / J_1 < 0.5 $.
Thus, it appears that the magnitude of frustration-induced quantum fluctuations 
are overestimated by spin-wave theory for the N\'eel state 
and severely underestimated for the stripe state.

\section{Summary and outlook}

\label{sec:outlook}

In this work we have studied the zero-temperature phase diagram of the
paradigmatic $J_1$-$J_2$ quantum Heisenberg model on a square lattice 
for arbitrary spin quantum number $S$ using an advanced 
implementation \cite{Tarasevych2022b} of the spin FRG approach proposed in Ref.~[\onlinecite{Krieg2019}]. Let us summarize the main technical advances reported in this work:

\begin{enumerate}

\item The innocent looking spin-length constraint $\bm{S}_i^2 = S ( S+1)$
implies an infinite tower of sum rules involving vertices of different order
which can be elegantly derived using our spin FRG formalism.

\item These sum rules can be used to construct
non-perturbative truncations of the
formally exact spin FRG flow equations.

\item    For frustrated magnets in reduced dimensions the implementation of these sum rules and the constraints imposed by spin conservation and ergodicity 
are crucial to construct sensible truncations of the  spin FRG flow equations.

\item In contrast to the pseudo-fermion FRG approach \cite{Reuther2010,Mueller2024}, the implementation of 
our spin FRG for frustrated magnets does not require heavy numerical calculations; the  results presented in this work have been  obtained in a few minutes
on a desktop  computer.
It is also not necessary to perform finite-size scaling to detect phase transitions between long-range ordered and paramagnetic phases in the spin FRG, unlike in the pseudo-fermion and pseudo-Majorana FRG implementations.
Which method is ultimately preferable for the study of frustrated quantum magnets of course also depends on their respective quantitative accuracy. In this respect, it would be desirable to compare the spin FRG with pseudo-fermion and pseudo-Majorana FRG in benchmark systems where exact results are known by other methods.

\item The spin FRG allows us to investigate quantum spin systems for arbitrary real values of the spin quantum number $S$,
without additional technical modifications or numerical costs. 
While implementations of the pseudo-fermion FRG for $ S > 1/2 $ exist \cite{Baez2017, Fukui2022, Mueller2024},
they come with additional technical complications because each spin-$S$ operator 
is represented by $ 2 S $ copies of spin-$1/2$ operators,
which are then expressed in terms of pseudo-fermions with an additional flavor index.
Due to this construction, they are also restricted to integer or half-integer values of $S$.

\end{enumerate}

The main physical result obtained in this work is the zero-temperature phase diagram of the  $J_1$-$J_2$ model as a function of $J_2 / J_1$ and the inverse spin quantum number $1/2S$ shown
in Fig.~\ref{fig4}. Although spin-wave theory produces a qualitatively similar
 phase diagram, the shape of the phase boundaries obtained by us
is more asymmetric and the extension of the paramagnetic phase significantly larger than predicted by spin-wave theory.

Our approach offers many promising opportunities for further research. The inclusion
of an external magnetic field is straightforward and does not add any significant 
complexity.
The door is now wide open to use our spin FRG approach to study 
ground state properties of
more complicated models  for frustrated magnets involving 
other lattices and less symmetric exchange couplings.

%
%
%

%
%
\begin{acknowledgments}
We thank Bj\"{o}rn Sbierski and Benedikt Schneider for useful discussions
and detailed comments on the manuscript. 
This work was financially supported by the Deutsche
Forschungsgemeinschaft (DFG, German Research Foundation) 
through Project No. 431190042.
\end{acknowledgments}

\appendix

\setcounter{equation}{0}

\renewcommand{\theequation}{A\arabic{equation}}

\renewcommand{\appendixname}{APPENDIX}

\renewcommand{\thesection}{\Alph{section}}

\section{Temperature rescaling} 

\label{app:temperature}

A significant advantage of the multiplicative deformation 
scheme \eqref{eq:deformation_scheme}
is that in the absence of external magnetic fields,
the temperature dependence can be removed from the problem entirely
by rescaling the deformation parameter and Matsubara frequencies as
\begin{subequations}
\begin{align}
\Lambda & = \bar{\Lambda} T \; , \;\;\; \;\;\; \bar{\Lambda} \in [ 0, 1 / T ] \; ,
\\[.1cm]
\omega & = \bar{\omega} T \; .
\end{align}
\end{subequations}
It is then straightforward to see from the generating functional \eqref{eq:vertex_expansion}
that the vertices should be rescaled as
\begin{widetext}
\begin{subequations}
\begin{align}
\Gamma_\Lambda^{ \alpha_1 \ldots \alpha_n \tilde{\alpha}_1 \ldots \tilde{\alpha}_m }
\left( 
\bm{k}_1 , \ldots , \bm{k}_n , 
( \bm{q}_1 , \omega_1 ) , \ldots , ( \bm{q}_m , \omega_m ) 
\right)
= & 
T^{ 1 - m }
\bar{ \Gamma }_{ \bar{\Lambda} = \Lambda / T }^{ \alpha_1 \ldots \alpha_n \tilde{\alpha}_1 \ldots \tilde{\alpha}_m }
\left( 
\bm{k}_1 , \ldots , \bm{k}_n , 
( \bm{q}_1 , \omega_1 / T ) , \ldots , ( \bm{q}_m , \omega_m / T ) 
\right)
\; ,
\\[.1cm]
\bar{ \Gamma }_{ \bar{\Lambda} }^{ \alpha_1 \ldots \alpha_n \tilde{\alpha}_1 \ldots \tilde{\alpha}_m }
\left( 
\bm{k}_1 , \ldots , \bm{k}_n , 
( \bm{q}_1 , \bar{\omega}_1 ) , \ldots , ( \bm{q}_m , \bar{\omega}_m ) 
\right)
= & 
T^{ m - 1 }
\Gamma_{ \Lambda  = \bar{\Lambda} T }^{ \alpha_1 \ldots \alpha_n \tilde{\alpha}_1 \ldots \tilde{\alpha}_m }
\left( 
\bm{k}_1 , \ldots , \bm{k}_n , 
( \bm{q}_1 , \bar{\omega}_1 T ) , \ldots , ( \bm{q}_m , \bar{\omega}_m T ) 
\right)
\; ,
\end{align}
\end{subequations}
\end{widetext}
where the rescaled $ \bar{\Gamma}_{\bar{\Lambda}} $ vertices do not depend 
explicitly on the temperature $T$.
In this way,
we obtain an effectively $T$-independent problem with $ T \to 1 $ at fixed $ \bar{\Lambda} $.
The $T$-dependence is then generated by flowing from $ \bar{\Lambda} = 0 $,
corresponding to $ T \to \infty $, to $ \bar{\Lambda} = 1 / T $.
In this form,
the only remaining dimensionful quantities are $ \bar{\Lambda} $ and $ J ( \bm{k} ) $,
which always appear as dimensionless product $ \bar{\Lambda} J ( \bm{k} ) $.
This rescaling also makes it obvious that by expanding the spin FRG flow equations in powers of $ \bar{\Lambda} J ( \bm{k} ) $ we obtain 
the high-temperature expansion
for the spin vertices.

\renewcommand{\theequation}{B\arabic{equation}}

\section{Spin-length sum rules from  spin FRG flow equations}

\label{app:sum_rules}

In order to efficiently derive the sum rules associated with the spin-length constraint \eqref{eq:length_constraint},
we write a general deformed Heisenberg Hamiltonian as
\begin{subequations}
\begin{align}
{\cal H}_\Lambda 
& = 
\frac{ 1 }{ 2 } \sum_{ i j } J_{ i j }^\Lambda \bm{S}_i \cdot \bm{S}_j 
\\[.1cm]
& = 
\frac{ 1 }{ 2 } \sum_{ i j } \left[ 
J_{ i j }^\Lambda + \delta_{ i j } C_\Lambda   - \delta_{ i j } C_\Lambda 
\right] \bm{S}_i \cdot \bm{S}_j 
\\[.1cm]
& = 
\frac{ 1 }{ 2 } \sum_{ i j } 
\left(  J_{ i j }^\Lambda + \delta_{ i j } C_\Lambda \right) \bm{S}_i \cdot \bm{S}_j 
- \frac{ N }{ 2 } C_\Lambda S ( S + 1 ) 
\; ,
\end{align}
\end{subequations}
with arbitrary, scale-dependent $ C_\Lambda $.
This rewriting leads to the following modifications in the flow equations:
\begin{enumerate}[label=(\roman*)]
\item
The constant energy shift adds an additional term
\begin{equation}
- \frac{ 1 }{ 2 } \partial_\Lambda C_\Lambda S ( S + 1 )
\end{equation}
to the right-hand side of the flow equation of the free energy density $ f_\Lambda $.
\item 
The on-site shift in the exchange coupling results in the replacement
\begin{equation}
\partial_\Lambda J_\Lambda ( \bm{q} ) \to 
\partial_\Lambda J_\Lambda ( \bm{q} ) + \partial_\Lambda C_\Lambda
\end{equation}
everywhere in all flow equations.
\item
The scale-dependent static spin susceptibility is now parametrized as
\begin{equation}
G_\Lambda ( \bm{k} ) = \frac{ 1 }{ 
J_\Lambda ( \bm{k} ) + C_\Lambda + \tilde{\Sigma}_\Lambda ( \bm{k} ) } \; .
\end{equation}
\end{enumerate}
The crucial point is now that we did not modify the system at all,
but only added a zero to the Hamiltonian regardless of the value of $ \Lambda $.
Hence,
all correlation functions must remain independent of $ C_\Lambda $.
Comparing with the original parametrization \eqref{eq:g_static} of the static spin susceptibility,
we immediately infer
\begin{equation}
\tilde{ \Sigma }_\Lambda ( \bm{k} ) = \Sigma_\Lambda ( \bm{k} ) - C_\Lambda \; ,
\end{equation}
where $ \Sigma_\Lambda ( \bm{k} ) $ is independent of $ C_\Lambda $.
As all other vertex functions in the hybrid functional are defined in terms
of spin correlation functions \cite{Tarasevych2021},
it follows that they must be independent of $ C_\Lambda $.
Consequently,
all $ \partial_\Lambda C_\Lambda $-contributions to their flow equations must cancel. 
For example,
the flow of the free energy density is given by
\begin{subequations}
\begin{align}
\partial_\Lambda f_\Lambda
= {} &
\frac{ 3 }{ 2 } \int_K \left[ \partial_\Lambda J_\Lambda ( \bm{k} ) \right] G_\Lambda ( K ) 
\;\;\; \text{at $C_\Lambda = 0$}
\\[.2cm]
= {} &
\frac{ 3 }{ 2 } \int_K \left[ \partial_\Lambda J_\Lambda ( \bm{k} ) + \partial_\Lambda C_\Lambda \right] G_\Lambda ( K )
\nonumber\\[.1cm]
&
- \frac{ 1 }{ 2 } \partial_\Lambda C_\Lambda S ( S + 1 )
\\[.2cm]
= {} &
\frac{ 3 }{ 2 } \int_K \left[ \partial_\Lambda J_\Lambda ( \bm{k} ) \right] G_\Lambda ( K )
\nonumber\\[.1cm]
&
+ \frac{ 3 }{ 2 } \left[ \partial_\Lambda C_\Lambda \right]
\left[ \int_K G_\Lambda ( K ) - \frac{ S ( S + 1 ) }{ 3 } \right] \; .
\end{align}
\end{subequations}
Comparing the first and last line of the above yields the susceptibility sum rule \eqref{eq:g_sum_rule}.
This scheme can be applied in the same way to each spin FRG flow equation
to yield an associated sum rule for the vertices.
In practice, 
this means that one can apply the rules (i)-(iv) given in Sec.~\ref{sec:sum_rules}
to any spin FRG flow equation to elegantly derive the associated sum rule,
since these rules generate precisely the $ \partial_\Lambda C_\Lambda $-contributions
that have to vanish in order to guarantee consistency of the correlation functions. 
Note also that because the hierarchy of spin FRG flow equations is infinite,
there is likewise an infinite tower of spin-length sum rules.
 

%
%
%
%
%
%
%

\renewcommand{\theequation}{C\arabic{equation}}

\section{Dynamics in the high-frequency approximation}

\label{app:high-frequency}

In this Appendix,
we give explicit expressions
for the flow equations (\ref{eq:u_sum_rule}-\ref{eq:pi_flow2})
in the high-frequency approximation \eqref{eq:Pi_via_A}.
They depend on the following four distinct Matsubara sums
\begin{widetext}
\begin{subequations}
\begin{align}
S_1 ( x ) 
= {} & 
\sum_{ \nu \neq 0 } \frac{ (\nu/T)^2 }{ \left[ (\nu/T)^2 + x \right]^2 }
=
\frac{ 
\sqrt{x} - \sinh\left( \sqrt{x} \right) 
}{ 
4 \sqrt{x} \left[ 1 - \cosh\left( \sqrt{x} \right) \right]
} 
\; ,
\\[.2cm]
S_2 ( x , y ) 
= {} & 
\sum_{ \nu \neq 0 } \frac{ (\nu/T)^4 }{ 
\left[ (\nu/T)^2 + x \right]^2 \left[ (\nu/T)^2 + y \right] }
\nonumber\\[.1cm]
= {} &
\frac{ 1 }{  8 ( x - y )^2 } \left[
2 \sqrt{x} ( x - 3 y ) \coth \left( \sqrt{x} / 2 \right) 
- x ( x - y ) \csch^2 \left( \sqrt{x} / 2 \right) 
+ 4 y^{ 3 / 2 }  \coth \left( \sqrt{y} / 2 \right)
\right]
\; ,
\\[.2cm]
S_3 ( x , y , z ) 
= {} & 
\lim_{ \omega \to \infty } \left[ \omega^2 \sum_{ \nu } 
\frac{ (\nu/T)^4 }{ \left[ (\nu/T)^2 + x \right]^2 }
\frac{ (\nu/T+\omega/T)^2 }{ (\nu/T+\omega/T)^2 + y }
\left(
1 + \frac{ z x }{ (\nu/T)^2 }
\right)
\left(
\frac{ y }{ (\nu/T+\omega/T)^2 } -
\frac{ z x }{ (\nu/T)^2 }
\right)
\right]
\nonumber\\[.1cm]
= {} &
\frac{ 1 }{ 8 } \left\{ 
4 \sqrt{y} \coth\left( \sqrt{y} / 2 \right)
- z \sqrt{x} \csch^2 \left( \sqrt{x} / 2 \right)
\left[
( z - 1 ) \sqrt{x} + ( z + 1 ) \sinh( \sqrt{x} )
\right]
\right\}
\; ,
\\[.2cm]
S_4 ( x )
= {} &
\sum_\nu \frac{ x^2 }{ \left[ (\nu/T)^2 + x \right]^2 }
=
\frac{ x + \sqrt{x}  \sinh( \sqrt{x} ) }{ 4 \left[ \cosh( \sqrt{x} ) - 1 \right] } 
\; .
\end{align}
\end{subequations}
Writing the leading coefficient $A_\Lambda ( \bm{k} )$
in the high-frequency expansion  \eqref{eq:Pi_via_A} of the polarization
$\Pi_{\Lambda} ( \bm{k} , \omega )$ in the form
\begin{equation}
A_\Lambda ( \bm{k} ) = T^2 G_\Lambda ( \bm{k} ) \Omega_\Lambda ( \bm{k} ) \; ,
\end{equation}
we find that the dynamic diagrams \eqref{eq:dynamic_a}
that contribute to the static self-energy are given by
\begin{equation}
T^2 G_\Lambda^2 ( \bm{q} ) 
\gamma_\Lambda ( \bm{k} , \bm{q} )
= 
2 \left[ \Sigma_\Lambda ( \bm{k} + \bm{q} ) - \Sigma_\Lambda ( \bm{k} ) \right]
S_1 \left( \Omega_\Lambda ( \bm{q} ) \right) 
- 2 G_\Lambda^{ - 1 } ( \bm{q} + \bm{k} ) 
S_2 \left( \Omega_\Lambda ( \bm{q} ) , \Omega_\Lambda ( \bm{q} + \bm{k} ) \right) \; .
\end{equation}
The polarization coefficient $A_{\Lambda} ( \bm{k} )$ 
itself satisfies the flow equation
\begin{subequations}
\begin{align}
\partial_\Lambda A_\Lambda ( \bm{k} )
= {} & 
U_\Lambda \dot{A}_\Lambda^{ (0) } ( \bm{k} )
+ \dot{A}_\Lambda^{ (1) } ( \bm{k} )
\; , \\[.1cm]
\dot{A}_\Lambda^{ (0) } ( \bm{k} )
= {} & 
T \dot{\Sigma}_\Lambda^{ ( 0 ) } ( \bm{k} ) \int_{ \bm{q} } \tilde{\gamma}_\Lambda^{ (A) } ( \bm{k} , \bm{q} )
\; , \\[.1cm]
\dot{A}_\Lambda^{ (1) } ( \bm{k} )
= {} & 
T \int_{ \bm{q} } \left[  
\partial_\Lambda J_\Lambda ( \bm{q} ) + \dot{\Sigma}_\Lambda^{ ( 1 ) } ( \bm{k} )
\right] 
\tilde{\gamma}_\Lambda^{ (A) } ( \bm{k} , \bm{q} )
\; , \\[.1cm]
\tilde{\gamma}_\Lambda^{ (A) } ( \bm{k} , \bm{q} )
= {} &
2 G_\Lambda^2 ( \bm{q} ) \left[ \Sigma_\Lambda ( \bm{q} + \bm{k} ) - \Sigma_\Lambda ( \bm{q} ) \right]
+ 2 G_\Lambda ( \bm{q} + \bm{k} ) 
S_3 \left( 
\Omega_\Lambda ( \bm{q} ) , 
\Omega_\Lambda ( \bm{q} + \bm{k} ) ,
G_\Lambda^{ - 1 } ( \bm{q} + \bm{k} ) G_\Lambda ( \bm{q} )
\right) 
\; .
\end{align}
\end{subequations}
Finally,
the static 4-point vertex in Eq.~\eqref{eq:u_sum_rule} reduces to
\begin{equation}
U_\Lambda = 
- \frac{
T^2 \int_{ \bm{k} } G_\Lambda^2 ( \bm{k} ) S_4 \left( \Omega_\Lambda ( \bm{k} ) \right)
\left[ \partial_\Lambda J_\Lambda ( \bm{k} ) + \dot{\Sigma}_\Lambda^{ (1) } ( \bm{k} ) \right] 
- \int_{ \bm{k} } S_1 \left( \Omega_\Lambda ( \bm{k} ) \right) \dot{A}_\Lambda^{ (1) } ( \bm{k} ) 
}{
T^2 \int_{ \bm{k} }G_\Lambda^2 ( \bm{k} ) S_4 \left( \Omega_\Lambda ( \bm{k} ) \right) \dot{\Sigma}_\Lambda^{ (0) } ( \bm{k} ) 
- \int_{ \bm{k} } S_1 \left( \Omega_\Lambda ( \bm{k} ) \right) \dot{A}_\Lambda^{ (0) } ( \bm{k} ) 
} \; .
\end{equation}
\end{widetext}


\begin{thebibliography}{99}
\bibitem{footnoteG}
Following Ref.~[\onlinecite{Tarasevych2021}]
we normalize the scale-dependent dynamic 
spin susceptibility $ G_\Lambda ( \bm{k} , \omega ) $
such that for free spins ($ \Lambda = 0 $),
it is given by
$ G_0 ( \bm{k} , \omega ) = 
\delta_{ \omega , 0 } G_0 ( \bm{k} ) =
\delta_{ \omega , 0 } / \Sigma_0 =
\delta_{ \omega , 0 } S ( S + 1 ) / 3 T $.
\bibitem{footnoteUV}
The expressions \eqref{eq:v_sum_rule} and \eqref{eq:u_sum_rule}
for the scale-dependent contributions $V_\Lambda ( \bm{k} )$ and $ U_\Lambda$
to the static 4-point vertex are by construction consistent
with the initial conditions at $  \Lambda = 0 $.
The reason is that the we initialize the flow with the exact vertices of isolated spins
that satisfy all sum rules automatically.
To see this explicitly in Eq.~\eqref{eq:u_sum_rule} requires Taylor-expansion 
of numerator and denominator to first order in $\Lambda$
and applying L'H\^{o}pital's rule.
\bibitem{footnoteclassical} 
With the explicit expressions \eqref{eq:dynamic}
for the contribution of the dynamic vertices,
it is obvious that they do not contribute in the classical limit 
$ S \to \infty $.
After the appropriate rescaling 
$ J_\Lambda ( \bm{k} ) S^2 \to J_\Lambda ( \bm{k} ) $,
$ \Sigma_\Lambda ( \bm{k} ) S^2 \to \Sigma_\Lambda ( \bm{k} )  $,
$ \Pi_\Lambda ( K )  S^{ - 2 } \to \Pi_\Lambda ( K ) $,
one finds that all dynamical diagrams contributing to
the 2-point flow equations (\ref{eq:u_sum_rule}-\ref{eq:pi_flow2}) are suppressed by a factor $ 1 / S^2 \to 0 $.
Hence, 
only the static problem remains in this limit,
and Eq.~\eqref{eq:u_sum_rule} for the quartic vertex simplifies to
$ U_\Lambda = 1 / [ T \int_{ \bm{q} } G_\Lambda^2 ( \bm{q} ) ] $.
\end{thebibliography}
\end{document}